\documentclass[twocolumn,showpacs,preprintnumbers,amsmath,amssymb]{revtex4}

\usepackage{graphicx}
\usepackage{dcolumn}
\usepackage{bm}

\begin{document}

\title{
Could the next generation of cosmology experiments exclude
supergravity ?}

\author{A. Barrau}
 \email{Aurelien.Barrau@cern.ch}
 \homepage{http://lpsc.in2p3.fr/ams/aurelien/aurel.html}
\affiliation{%
Laboratory for Subatomic Physics and Cosmology\\
CNRS-IN2P3 / Joseph Fourier University\\
53, av des Martyrs, 38026 Grenoble cedex, France 
}%

\author{N. Ponthieu}
 \email{Nicolas.ponthieu@ias.u-psud.fr}
\affiliation{%
Institute for Space Astrophysics\\
CNRS-INSUE / Paris-Sud University\\
91405, Orsay, France
}%

\date{\today}

\begin{abstract}
Gravitinos are expected to be produced in any local supersymmetric
model. Using their abundance prediction as a function of the reheating
energy scale, it is argued that the next generation of {\it Cosmic
Microwave Background} experiments
could exclude supergravity or strongly favor "thermal-like" inflation
models if $B$ mode polarized radiation were detected. Galactic cosmic--ray
production by evaporating primordial black
holes is also investigated as a way of constraining
the Hubble mass at the end of inflation. Subsequent limits on
the gravitino mass and on the related grand unification parameters are
derived.\\

\begin{center}
Phys. Rev. D 69 (2004) 105021
\end{center}

\end{abstract}

\maketitle

\section{Introduction: gravitinos in the early universe}

Although not yet experimentally discovered, supersymmetry (SUSY) is still the
best - if not the only - natural extension of the standard model
of particle physics. It could provide a general framework to understand the
origin of the fundamental difference between fermions and bosons and
could help to resolve the difficult problem of mass hierarchies,
namely the instability of the electroweak scale with respect to
radiative corrections. In global supersymmetry, the generator spinors
$\xi$ are assumed to obey $\partial_{\mu}\xi=0$ \cite{freedman}. If
one wants to deal with local supersymmetry, or supergravity, this
condition must be relaxed and $\xi$ becomes a function of the space coordinates
$x$. New terms, proportional to $\partial_{\mu}\xi(x)$, must be
canceled by introducing a spin 3/2 particle, called gravitino, as
vector bosons are introduced in gauge theories. The gravitino is part
of an N=1 multiplet which contains the spin 2 graviton (see
\cite{olive} for an introductive review) and, in the broken phase of
supergravity, super-Higgs effects make it massive through the
absorption of the Nambu-Goldstone fermion associated with the SUSY
breaking sector.\\

It has long been known that if the gravitino is unstable some severe
constraints on its mass must be considered in order to avoid entropy
overproduction \cite{weinberg}: $m_{3/2}\gtrsim 10$ TeV. On the other hand,
if the gravitino is stable, its mass should satisfy $m_{3/2}\lesssim 1$~keV \cite{pagels}
to keep the gravitinos density smaller than the full Universe density
($\Omega_{3/2}<\Omega_{tot}$).  In spite of the huge
dilution, those constraints are not fully evaded by inflation as
gravitinos should be reproduced by scattering processes off the
thermal radiation after the Universe has reheated
\cite{nano,khlo,ellis,Leigh,Fujisaki,ellis2,Kallosh}. 
As the number density of such secondary gravitinos is expected to be
proportional to the reheating temperature, it is possible to relate
the energy scale of inflation with the requirement that they are not
overproduced.\\

In the first part of this paper, the next generation of 
cosmic microwave background (CMB) detection experiments is considered as a way of possibly
excluding supergravity. It is shown that the energy scale of inflation required to
produce an observable amount of tensor mode in the background radiation is not compatible with
local supersymmetry in the standard cosmological scenario. In the second part,
a new way of constraining the gravitino mass, based on evaporating
primordial black holes, is investigated. Taking into account that the black hole masses cannot be
much smaller than the Hubble mass at the formation epoch, it is suggested that
 a detection of
cosmic--rays produced by the Hawking mechanism would lead to a lower bound on the
reheating scale and, therefore, on the gravitino mass. Links with grand-unified
models are given, as an example, in the conclusion. Finally, 
the basics of the
propagation model used to relate the source term to the local spectrum are given
in the Appendix A.

\section{Tensor mode in the cosmological background}

Observational cosmology has recently entered a new era thanks to
several experiments dedicated to the CMB
measurements 
(see \footnote{http://www-dapnia.cea.fr/Phys/Sap/Activites/Science/
Cosmologie/Fond/page.shtml\#exp}), {\it e.g.} Maxima, BOOMERanG, ACBAR,
DASI, CBI, VSA, A{\small RCHEOPS}, and WMAP. They give strong evidences
in favor of the inflationary scenario: a density extremely close to
the critical value, a nearly scale invariant power spectrum, and a
gaussian structure of the perturbations. Furthermore, in addition to
the temperature anisotropies, the polarization of the CMB has also
been recently observed \cite{dasi,map}. For the time being, only the
even-parity $E$ mode has been detected and the odd-parity $B$ mode is still to
be discovered. The latter is of specific importance as it would
probe the primordial gravitational waves through tensor
perturbations. Their amplitude can be expressed with the Hubble
parameter and the potential of the scalar field driving inflation
\cite{kinney}: $$T=\left( \frac{H}{2\pi M_{Pl}} \right) 
= \frac{2V(\phi)}{3\pi M_{Pl}^4}$$

where $M_{Pl}=(8\pi G)^{-1/2}=2.4\times 10^{18}$~GeV is the Planck
mass. The important point is that the tensor/scalar ratio $r =
6.9M_{Pl}^2(V^\prime/V)^2$ can be related to the energy scale of
inflation $E_{infl}$ \cite{lyth} by

$$E_{infl}\approx \left( \frac{r}{0.7}
\right)^{1/4}\times 1.8 \times 10^{16}~{\rm GeV}.$$

The amplitude of the polarization $B$ mode is therefore directly
proportional to $E_{infl}$.\\

\begin{figure}
\scalebox{0.55}{\includegraphics{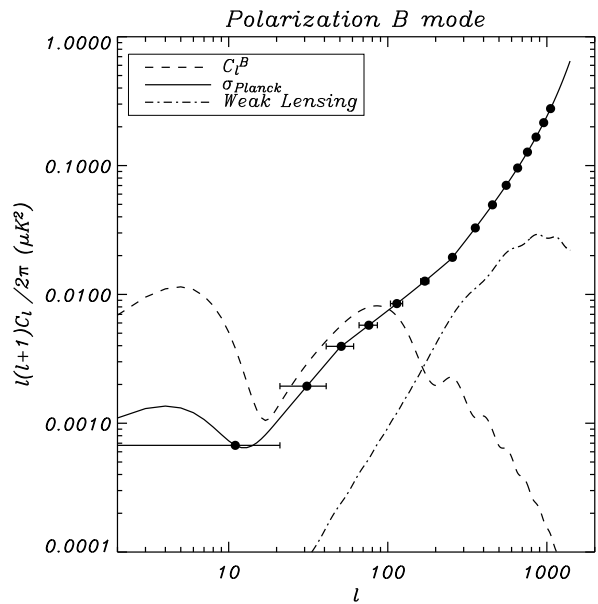}}
\caption{\label{fig:planck_sensiv.ps} 
Sensitivity (1~$\sigma$) to polarization of the Planck satellite
(solid line) versus the expected $B$ mode polarization in a standard
$\Lambda$CDM cosmology with an inflation energy scale of $\sim 10^{16}$
GeV (dotted line). Planck should provide significant detection of
this tensor mode, especially at low multipole $\ell$ where reionization
boosts the power spectrum. The $B$ mode induced by weak lensing is
also represented (dot--dash line) and dominated the primordial
spectrum for $\ell \leq 200$.}
\end{figure}

Figure \ref{fig:planck_sensiv.ps} shows the 1~$\sigma$ sensitivity of
the Planck satellite to polarization, as computed with CMBfast \footnote{http://www.cmbfast.org}. 
On the same plot, 
the $B$ mode polarization in a standard $\Lambda$CDM cosmology with an
inflation energy scale  $E_{infl} \sim 10^{16} {\rm GeV}$ (dotted
line) is also represented. Increasing (resp. lowering) $E_{infl}$ would result in
increasing (lowering) the amplitude of the primordial $B$ mode thus
making it easier (more difficult if not impossible) to detect. On the
contrary, the level of the expected $B$ mode induced by weak lensing
is fixed and rather accurately predicted since it results from lensing
effects on the polarization $E$ mode  due to scalar perturbations that
are now well constrained. The challenge in the detection of the
primordial $B$ mode and the estimation of $E_{infl}$ is then to have a
sensitive enough experiment and to avoid contamination by weak
lensing. For the Planck experiment, the major hope is the detection at low
$\ell$ thanks to the high reionization optical depth suggested by WMAP
\cite{page}. In the case of limited sky coverage experiments, the weak
lensing contribution will have to be removed.\\

With the Planck sensitivity, the $B$ mode should be
detected ($3\sigma$) if $E_{infl} > 10^{16}$~GeV \cite{jaffe,
nico}. This case would be in severe conflict with most
supersymmetric models. Indeed, in mSUGRA, the gravitino mass is, by
construction, expected to lie around the electroweak scale, {\it i.e.},
in the 100~GeV - 1~TeV range
\cite{djouadi}. Considering that Deuterium and $^{3}$He should not be
overproduced by photodissociation of $^{4}$He below 700~GeV and that
Deuterium should not be destroyed beyond the allowed observational values
\cite{walker} above 700~GeV \cite{kawa}, the reheating temperature
must remain lower than $2\times 10^9~{\rm GeV}$ if the branching ratio
of gravitinos into photons and photinos is assumed to be unity and lower
than $5\times 10^{11}~{\rm GeV}$ with a conservative branching ratio
of 1/10. The large difference between those limits and the energy
scale required to produce a measurable amount $B$ mode polarization
makes the exact value of the branching ratio of gravitinos into
photons and photinos irrelevant. A detection of the polarization 
$B$ mode
by the Planck satellite would therefore disfavor mSUGRA in
{\it standard} cosmology.\\

In gauge-mediated SUSY breaking alternative scenarios, mostly
interesting in accounting for a natural suppression of the rate of
flavor-changing neutral-current due to the low energy scale, the
situation is even more constrained. In this case, gravitinos are the
lightest supersymmetric particles and requiring their density not
exceed the total density imposes an upper limit
on $T_{RH}$ between $10^6$ and $10^3$~GeV for masses between 10
MeV and 100 keV \cite{moroi}. Although some refined models can relax
those constraints \cite{fujii}, local supergravity would, in this case also,
be in serious
trouble if the reheating temperature was high enough to be probed by
the Planck experiment.\\

A possible way to get around these conclusions is to assume that a
substantial amount of entropy was released after the gravitinos and
moduli production, that would dilute them according to the entropy
conservation ($n/s \simeq $ cte). Such a scenario can be realized while
keeping the inflationary scale high, {\it e.g.}, in thermal inflation
\cite{lyth1,lyth2}. 
Some studies \cite{asaka} even show that a wide modulus mass region
($m_{\Phi}\approx10$ eV - $10^4$ GeV) would be allowed but it requires
in most cases a very small reheating temperature. Recently, the
curvaton scenario \cite{wands} has also attracted a considerable interest
as it generates a huge amount of entropy through a scalar field that
dominates the radiation at a given epoch. One can then argue that a
detection of tensor mode polarization, would strongly favor ``thermal
like'' inflation scenarios if supergravity is to remain as the preferred
extension of the standard model of particle physics. Interestingly,
if evidence in favor of local supsersymmetry were obtained either by
colliders or by independent astroparticle experiments, this could even be
a very promising observational signature for thermal inflation.\\

\begin{figure}
\scalebox{0.45}{\includegraphics{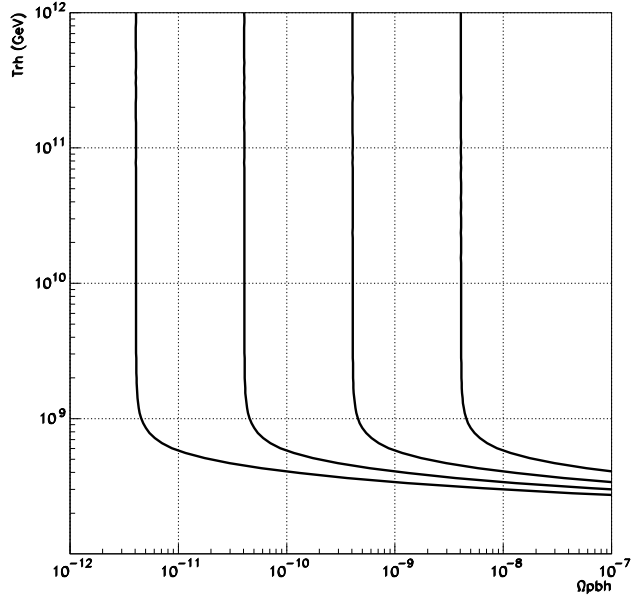}}
\caption{\label{reheat} Possible reheating temperatures $T_{RH}$ as a function of the PBH density (normalized to
the critical density) for different antideuteron flux at 100 MeV~: $2\times
10^{-7}$, $2\times 10^{-8}$, $2\times 10^{-9}$, $2\times 10^{-10}$ from 
right to left in ${\rm
m}^{-2}{\rm s}^{-1}{\rm sr}^{-1}{\rm GeV}^{-1}$}
\end{figure}

Fortunately, the Planck satellite is not expected to be the ultimate
experiment to study the CMB polarization and several improvements can
be expected in the future. However, as pointed out in
Refs.~\cite{knox,hirata}, there remains a lower limit to the removal
of the  polarization $B$ mode foreground induced by gravitational
lensing which sets at present time the lower limit on the detectable
inflation scale to a few times $10^{15}$~GeV. This scale remains,
however, particularly interesting if the fundamental scalars driving
the phenomenon are related with grand unification
since it lies around the GUT energy
(between $10^{15}$~GeV and $3\times 10^{16}$~GeV depending on whether
supersymmetry is considered or not). It therefore makes
sense to improve the polarization sensitivity to reach the capability
to probe the typical GUT scale where inflation could have occurred if
the gravitino limit is ignored.

\begin{figure}
\scalebox{0.45}{\includegraphics{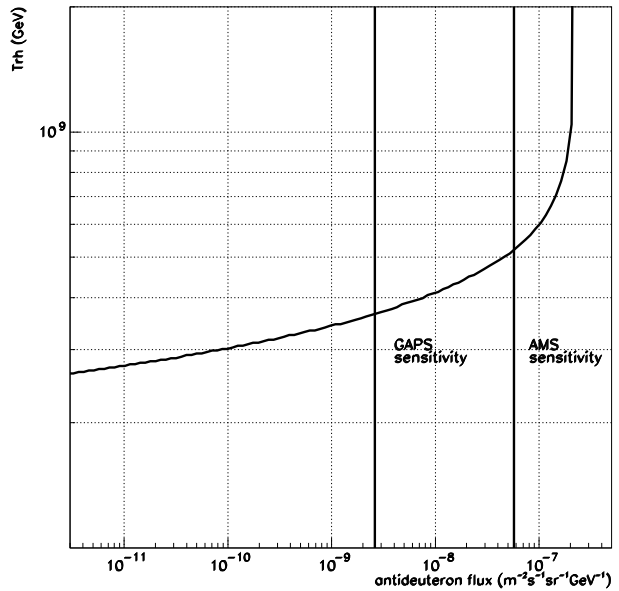}}
\caption{\label{plot} Lower limit on the reheating temperature $T_{RH}$ as a function of the 100
MeV antideuteron flux. }
\end{figure}

\section{cosmic--rays from evaporating black holes}

Another interesting way to experimentally probe the reheating
temperature would be to look for evaporating primordial black holes
(PBHs). Such black holes should have formed in the early Universe if
the density contrast was high enough on small scales.  Many different
possible scenarios have been suggested to allow for an important PBH
density~(see \cite{khlobook} for a review): a dust-like stage
\cite{khlopbh}, general first order phase transitions \cite{karsten},
a scale in the power spectrum \cite{po,barpo}, to mention only the
currently most discussed possibilities. Such PBHs of mass $M$ should
evaporate, following a Plank-like spectrum with temperature
$T=hc^3/(16\pi kGM)$, which was derived by Hawking \cite{hawking}
using the usual quantum mechanical wave equation for a collapsing
object with a postcollapse classical curved metric instead of a
precollapse Minkowsky one. If those black holes are present in our
galaxy (even with densities as low as $\Omega_{PBH}\sim 10^{-9}$), the
emitted quanta should contribute to the observed cosmic--rays.  Among
them, two kinds of particles are especially interesting: antiprotons
and gamma--rays. Antiprotons are useful because the astrophysical
background coming from spallation of cosmic--rays on the interstellar
medium (so-called secondary particles) is very small (the ratio
$\bar{p}/p$ is smaller than $10^{-4}$ whatever the considered energy)
and very well known \cite{fio}. A tiny excess due to evaporating black
holes could therefore be easily probed in the low energy range
\cite{barrau1} since the shape of the PBH spectrum is dominated by
fragmentation processes and is then softer than the secondary
spectrum. Gamma--rays, coming both from direct emission and from the
decay of neutral pions, take advantage of the very small optical depth
of the Universe for $\sim 100$ MeV radiation \cite{carr}: the source
emission can be probed up to redshifts $z\sim 700$. Furthermore, the
signal to noise ratio is optimal at this energy as the PBH spectrum
becomes softer ($dN/dE\propto E^{-1} \rightarrow dN/dE\propto E^{-3}$)
above 100 MeV (roughly corresponding to the QCD confinement scale)
because of partons hadronization and integrated redshift effects
\cite{janeapj}. \\

Using those cosmic-rays, the experiments are currently sensitive to
PBHs with masses between $10^{12}$ and $10^{14}$~g.  Those values
can be intuitively understood as resulting from two opposite
effects. On the one hand, the temperature favors the light ({\it i.e.}
hot) black holes but their number density is very small~: by
integrating the Hawking flux over energy, it is straightforward to
show that the mass spectrum {\it must} be proportional to $M^2$ below
$M_*=5\times 10^{14}$~g (the initial mass of a black hole whose
lifetime is equal to the age of the Universe) whatever the details of
the formation mechanism \cite{barrau3}. This is mostly due to the fact
that the low-mass behavior is fully governed by the evaporation
process, as obtained by writing $dn/dM=(dn/dM_i)\times (dM_i/dM)$
where $M$ stands for the current mass value and $M_i$ for the initial one. The
evolution term $dM_i/dM$ is simply determined from $M_i\approx(3\alpha
t + M^3)^{1/3}$ where $\alpha
\approx \{ 7.8d_{s=1/2}+3.1d_{s=1} \} \times 10^{24}$
g$^3$s$^{-1}$ accounts for the number of available degrees of freedom
with $d_{s=1/2}=90$ and $d_{s=1}=27$ in the standard model
\cite{MacGibbon3}.  On the other hand, the "number density" effect
favors the heavy black holes but their low temperature makes the
emission rate very small, especially when heavy hadrons are considered.\\

The important point for this study is that only black holes formed
after inflation would contribute to the observed phenomena as those
formed before were exponentially diluted.  Furthermore, whatever the
considered formation mechanism, either through the usual collapse of
high density-contrast primordial gaussian fluctuations or for near
critical phenomena \cite{yoko}, the PBH mass at the formation epoch is
close to the horizon mass at the same time. It cannot be larger as the
considered points would not be in causal contact and it cannot be much
smaller as they would, in this case, more probably have formed before
(as taken into account in the usual Press-Schechter formalism). It
means that if the evaporation process was detected, the Hubble mass at
the reheating time should be small enough not to induce a cutoff in
the PBH mass spectrum which would make the light black holes abundance
totally negligible. The best upper limit available on the density of
PBHs around $M_*=5\times 10^{14}$~g, taking into account both the
details of the source term evolution and the background from galaxies
and quasars, is currently: $\Omega_{PBH}(M_*)<3.3\times 10^{-9}$
\cite{barrau2}.\\

Fortunately, some hope for future detection is still possible thanks
to antideuterons: those nuclei are expected to be very rarely formed
by spallation processes below a few GeV for kinematical reasons.  The
threshold for an antideuteron production is $E=17\, m_p$ (total
energy) in the laboratory, 2.4 times higher than for antiproton
production.  The center of mass is, therefore, moving fast and it is
very unlikely to produce an antideuteron nearly at rest (in the 100
MeV - 1 GeV range) in the laboratory.  On the other hand, they could
be emitted in this energy range by evaporating PBHs and could be
probed by the new generation of cosmic--ray detectors: the AMS
experiment \cite{barrau4} and the GAPS project \cite{gaps}.  To obtain
this result, a coalescence model (see \cite{csernai} for a review) was
used, based mainly on phase space considerations: the antideuteron
density in momentum space is proportional to the product of the proton
density with the probability of finding a neutron within a small
sphere of radius $p_0$ around the proton momentum. Thus: $$
\gamma\frac{d^3N_d}{dk^3_d}=\frac{4\pi}{3}p_0^3\left(\gamma
\frac{d^3N_p}{dk^3_p}\right)\left(\gamma \frac{d^3N_n}{dk^3_n}\right)
$$
where $p_0$ is the coalescence momentum whose uncertainty window is of the order
of 60-280 Mev in extreme cases.
The Hawking
spectrum has then been convolved with the fragmentation functions, as
obtained with the PYTHIA \cite{pythia} Monte-Carlo simulation of the
Lund model: 
$$
\frac{{\rm d}^2N_{\bar{D}}}{{\rm d}E{\rm d}t}=
\sum_j\int_{Q=E}^{\infty}\alpha_j\frac{\Gamma_{s_j}(Q,T)}{h}
\left(e^{\frac{Q}{kT}}-(-1)^{2s_j}\right)^{-1}$$
$$
\times\frac{{\rm d}g_{j\bar{D}}(Q,E,p_0)}{{\rm d}E}{\rm
d}Q
$$
where ${\rm d}g_{j\bar{D}}(Q,E,p_0)/{\rm d}E$ is the number of
antideuterons formed with an energy between $E$ and $E+dE$ by a partonic
jet of type $j$ and energy $Q$,
evaluated with the coalescence model for a given
momentum $p_0$, $\alpha_j$ is the number of degrees of freedom, $s$ is the
spin, and
$\Gamma_s$ is the absorption probability. This coalescence condition (finding 
an antiproton and an
antineutron within the same jet with a momentum difference smaller than $p_0$)
was directly tested in the $\bar{p}-\bar{n}$ center of mass frame as $p_0$
is not Lorentz-invariant and implemented within the PYTHIA simulation.
This individual flux in then convolved with the PBH mass spectrum.
To obtain the {\it top of the atmosphere} (experimentally measurable) 
spectrum, the emitted antideuterons have been propagated
within the Galaxy using the  diffusion model of
\cite{fio}, briefly recalled in 
the appendix at the end of this paper. Finally, the resulting flux
were solar-modulated in the force-field approximation.\\

Figure \ref{reheat} shows the possible values of the reheating
temperature as a function of the density of PBHs at $5\times10^{14}$~g
for different PBH-induced antideuteron flux at 100 MeV (ranging from
$2\times 10^{-7}~{\rm m}^{-2}{\rm s}^{-1}{\rm sr}^{-1}{\rm GeV}^{-1}$,
the maximum value consistent with the gamma--ray upper limit, down
to $2\times 10^{-10}~{\rm m}^{-2}{\rm s}^{-1}{\rm sr}^{-1}{\rm
GeV}^{-1}$). They were obtained with conservative values of
all the free parameters entering the model, astrophysical quantities
being totally bounded by an exhaustive study of the heavy nuclei data
\cite{david}. As expected, there is a degeneracy between the $\bar{D}$
flux and $\Omega_{PBH}$: the same amount of particles can be produced
either by a high normalization of the black hole spectrum and a cutoff
in the high mass range ({\it i.e.} a low reheating temperature value)
of by a low normalization of the black hole spectrum and a cutoff in
the low mass range ({\it i.e.} a high reheating temperature value).
This means that, in the case of detection, it should be possible to give a
lower limit on the reheating temperature. Of course, the larger the
antideuteron flux, the better the constraint on $T_{RH}$. As shown on
this figure, for a fixed value of the flux, whatever the value of
$\Omega_{PBH}$, the reheating temperature value cannot be arbitrarily
low since the mass spectrum cannot be cut much above masses roughly
corresponding to temperatures of the order of the $\bar{D}$ mass ({\it
i.e.} $T_{BH}\sim$ a few GeV and $T_{RH}\sim$ a few $10^8$ GeV). The
other way round, whatever the value of $T_{RH}$, the density of black
holes cannot be arbitrarily low since even without any cutoff the
source term must remain high enough to account for the considered
flux. Naturally, this approach assumes that the measured antideuterons
are indeed produced by evaporating black holes. The only other serious
candidate as a source of light antinuclei in the low energy range are
annihilating supersymmetric particles. It has been demonstrated
\cite{fio2} that only neutralinos with masses around 100-200 GeV could
contribute to the observed antideuteron flux. As this mass range will
be probed by the Large Hadron Collider, it should be possible to
distinguish between antideuterons induced by PBHs and by SUSY
particles (some reconstruction problems could occur if the mass
spectrum is strongly degenerated, especially between the lightest
neutralinos and charginos, but this would hide the lightest
supersymmetric particles only for masses in the TeV range).\\

\begin{figure}
\scalebox{0.45}{\includegraphics{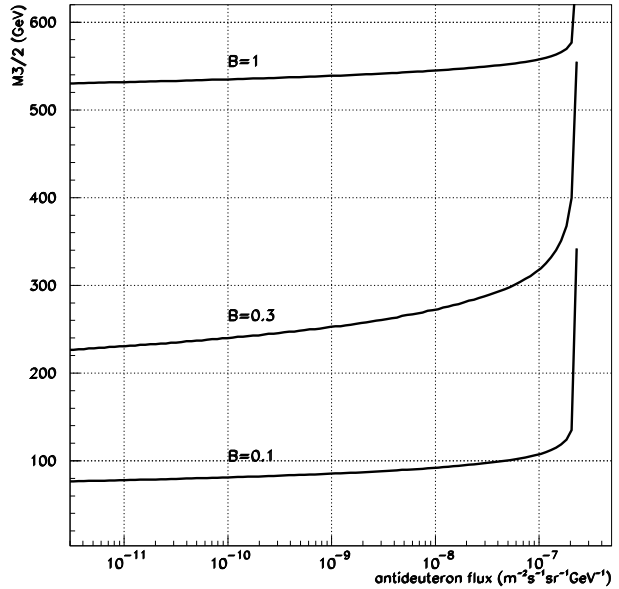}}
\caption{\label{gravitino} Lower limit on the gravitino mass as a function of
the measured antideuteron flux for three different branching ratios.}
\end{figure}

In the case where they are indeed coming from black holes, Fig.~\ref{plot} gives
the reheating temperature value as a function of the  measured
$\bar{D}$ flux.  This result was obtained by 
varying values of the 100 MeV
antideuteron spectrum combined with the upper limit coming from
\cite{barrau2} and \cite{barrau1} ($\Omega_{PBH}<3\times 10^{-9}$) for the
corresponding reheating scale (evaluated by the previously given method). As
expected, the limit becomes more stringent when the measured flux is
higher and diverges when it goes to the maximum allowed
value (otherwise it contradicts previously given limits).  When
compared with the upper bound coming from Big--Bang nucleosynthesis,
this translates into a lower limit on the gravitino mass $m_{3/2}$.
This can be derived by solving the Boltzmann equation for the
gravitino number density $n_{3/2}$ \cite{kawa}:
$$\frac{dn_{3/2}}{dt}+3Hn_{3/2}=
<\Sigma_{tot}v_{rel}>n_{rad}^2-\frac{m_{3/2}}{<E_{3/2}>}
\frac{n_{3/2}}{\tau_{3/2}}$$
where $H$ is the Hubble parameter, $n_{rad}=\zeta(3)T^3/\pi^2$ is the
number density of the scalar bosons in thermal bath, $v_{rel}$ is the
relative velocity of the scattering radiation, $m_{3/2}/(<E_{3/2}>)$
is the averaged Lorentz factor, $\tau_{3/2}$ is the lifetime of the
gravitino (computed from the supergravity lagrangian \cite{crem}) and
$\Sigma_{tot}$ is the total cross section (computed in the MSSM
framework). Gravitinos are then assumed to decay mostly into photinos
and photons, whose pair scattering off the background radiation,
photon-photon scattering, pair creation on nuclei, compton scattering,
inverse compton scattering of $e^+/e^-$ and induced leptonic cascades
are taken into account. Requiring that the subsequent
photo-dissociation of light elements does not modify the {\it Big Bang
Nucleosynthesis} scenario beyond experimental constraints, the upper
limit of the reheating temperature can be numerically computed as a
function of the gravitino mass \cite{kawa}.  Figure \ref{gravitino}
gives this bound as a function of the  measured antideuteron
flux at 100 MeV for three different branching ratios $B$ of gravitinos
into photons and photinos ranging from 0.1 (lowest curve) to 1 (upper
curve). As the reheating temperature lower limit is extremely
sensitive to the gravitino mass in the 100 GeV - 1 TeV range
\cite{kawa}, the curves are quite flat, except when the required value
of $T_{RH}$ enters the diverging region.  Although the accurate value
of $B$ is model dependent, it can safely be taken as lying in the 0.1-1
range, as usually assumed in most studies.  Once again, if a "thermal-like"
inflation phase occurred, those limits do not stand anymore but could
lead to important indications in favor of such a scenario if the
gravitino was independently shown to be lighter than those values.\\

It is important to notice that a great amount of work has also been
recently devoted to the non-thermal production of gravitinos and
moduli fields (dilaton and modulus fields appearing in the framework
of superstring theories which acquire mass through the nonperturbative
effects of the supersymmetry breaking). Most papers claim that the
upper limit on the reheating temperature must be drastically decreased
(by up to 7 orders of magnitude \cite{Giudice}). Those results being
still controversial, they were not taken into account in this work but
they can only reinforce our conclusions and improve our limits.

\begin{figure}
\scalebox{0.45}{\includegraphics{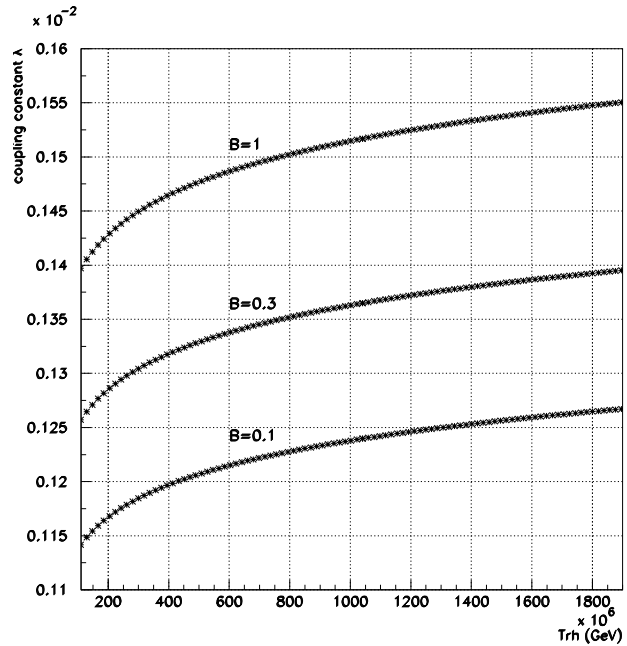}}
\caption{\label{lambda} Lower limit on the coupling constant $\lambda$ as a function
of the reheating temperature.}
\end{figure}

\section{prospects and conclusion}

It must be pointed out that such possible constraints
on the gravitino mass can be translated into constraints on more
fundamental parameters, making them very valuable in the search for
the allowed parameter space in {\it grand unified} models. As an
example, in models leading naturally to mass scales in the
$10^2$-$10^3$~GeV range through a specific dilaton vacuum
configuration in supergravity, the gravitino mass can be related with
the GUT parameters \cite{tkach}:
$$
m_{3/2}=\left(
\frac{5\pi^{\frac{1}{2}}\lambda}{2^{\frac{3}{2}}}\right)^{\sqrt{3}}
(\alpha_{GUT})\left(\frac{M_{GUT}}{M_{Pl}}\right)^{3\sqrt{3}}M_{Pl}.
$$
With $M_{GUT}\sim 10^{16}$~GeV and a gauge coupling $\alpha_{GUT}\sim
1/26$. 
The superpotential value in the  dilaton direction defines the magnitude
of the coupling constant $\lambda$ of the self-interacting 24 multiplet.
Figure~\ref{lambda} shows how the lower value on $\lambda$
 evolves as a function of the reheating temperature
which could be probed by the previously given method, for three different
branching ratios. Although not
very constraining, this lower limit of the order $1.4\times10^{-3}$ over
the full tested range for $B=1$ could be one of the first experimental constraints on
$\lambda$.

The next generation of CMB experiments will face a new
situation. Important efforts are devoted to the search for the
polarization $B$ mode \cite{minneapolis} and the sensitivity should reach
scales of inflation of order $10^{15} - 10^{16}$~GeV. This value is slightly
higher than the GUT scale if supersymmetry is ignored ({\it i.e} if
gravitinos production is expected not to have occurred), and slightly lower than
the GUT scale if supersymmetry is taken into account ({\it i.e.} in the case
gravitinos are expected to be produced by scattering processes). 
Considering that the grand
unified scale is the highest natural value for the reheating temperature, this 
means that, 
if a significant amount of entropy was not released after the moduli production,
it should not be possible to detect those tensor modes in both scenarios.

On the other hand, cosmic--ray experiments could be sensitive enough to
investigate the allowed reheating temperatures if small black holes were
formed at the end of inflation. In this case, important limits could be
derived on the gravitino mass and on the related GUT parameters.\\

\section{Appendix : Antideuteron Flux Computation}

In this two-zone approach, the geometry of the Milky Way is a
cylindrical box whose radial extension is $R=20$ kpc from the galactic
center, with a matter (stars) disk whose thickness is $2h=200$ pc and
a diffusion halo whose extent is the major source of uncertainty
(taken into account in the analysis).  The five parameters used are
$K_0$, $\delta$ (describing the diffusion coefficient $K(E)=K_0 \beta
R^{\delta}$), the halo half height L, the convective velocity $V_c$
and the Alfv\'en velocity $V_a$. They are varied within a given range
determined by an exhaustive and systematic study of cosmic ray nuclei
data \cite{david}. The same parameters as employed to study the
antiproton flux \cite{barrau1} are used again in this analysis.  The
antideuterons density produced by evaporating PBHs per energy bin
$\psi_{\bar D}$ obeys the following diffusion equation: $$
\left\{V_c\frac{\partial}{\partial z} -K\left(\frac{\partial^2}{{\partial
z}^2}\left(r\frac{\partial}{\partial z}\right)\right)\right\}\psi_{\bar
D}(r,z,E) +$$
$$
2h\delta (z) \Gamma_{\bar D}\psi_{\bar D}(r,0,E)= 
q^{prim}(r,z,E)
\label{eqdiff}
$$
where $q^{prim}(r,z,E)$ corresponds to the source term.
The total collision rate is given by
$\Gamma_{\bar D} = n_H \sigma_{\bar D H}v_{\bar D}$ where 
$\sigma_{\bar D H}$ is
the total antideuteron cross-section with protons and
the hydrogen density, assumed to be constant all over the disk, has 
been fixed to
$n_H=1$ cm$^{-3}$.

Performing Bessel transforms, all the quantities can be expanded over the
orthogonal set of Bessel functions of zeroth order:
$$
\psi_{\bar D} = \sum_{i=1}^{\infty}N_i^{\bar D,prim}
J_0(\zeta_i (x))
$$
and the
solution of the equation for antideuterons can be written as
$$
N_i^{\bar D,prim}(0)=\exp\left(\frac{-V_c
L}{2K}\right)\frac{y_i(L)}{A_i\sinh\left(S_iL/2\right)}
$$
where
\begin{displaymath}
\left\{
\begin{array}{l}
y_i = 2\int_0^L \exp\left(\frac{V_c}{2K}(L-z')\right)\sinh\left(\frac{S_i}{2}(L-z')\right)q_i^{prim}(z')
dz'\\
S_{i} \equiv \left\{
{\displaystyle \frac{V_{c}^{2}}{K^{2}}} \, + \,
4 {\displaystyle \frac{\zeta_{i}^{2}}{R^{2}}}
\right\}^{1/2}\\
A_{i} \equiv 2 \, h \, \Gamma^{ine}_{\bar{D}}
\; + \; V_{c} \; + \; K \, S_{i} \,
{\rm coth} \left\{ {\displaystyle \frac{S_{i} L}{2}} \right\}
\;\;
\end{array}
\right.
\end{displaymath}

In this model, energy changes (predominantly ionization losses, adiabatic losses and
diffusive reacceleration) are taken into account via a second order
differential equation for $N_i^{\bar{D},prim}$.  The spatial
distribution $f(r,z)$ of PBHs was assumed to follow
$$
f(r,z)=\frac{R_c^2 + R_\odot^2}{R_c^2 +  r^2 + z^2}
\label{halo}
$$ where the core radius $R_c$ has been fixed to 3.5 kpc and $R_\odot$=8 kpc.
This profile corresponds to the isothermal case with a spherical symmetry,
the uncertainties on $R_c$ and the consequences of a possible flatness
have been shown to be irrelevant in \cite{barrau1}.

\vspace{2cm}

{\it Acknowledgments:} the authors would like to thank P. Salati, D. Maurin, R.
Taillet and F. Donato who developed the propagation model used in this work. We
also would like to thank A. Lucotte for very valuable informations on the neutralino
detection. 


\bibliography{article}
      
\end{document}